\documentclass[useAMS,usenatbib]{mn2e}

\usepackage{epsfig,natbib,color}
\usepackage{myaasmacros,amssymb}
\usepackage{graphicx}
\usepackage{placeins}
\usepackage{times}
\newcommand{\dd}{\mathrm{d}}
\newcommand{\angst}{\mathrm{\AA}}

\def \siii {{\rm Si\,{\sc ii}}}

\def \hi {{\,\rm H\,{\sc i}}}
\def \hii {{\,\rm H\,{\sc ii}}}
\def \cm {{\,\rm cm}}
\def \Mpc{{\,\rm Mpc}}
\def \kms{{\,\rm km\textrm{ }s^{-1}}}
\def \kpc{\,\mathrm{kpc}}
\def \pc{{\,\rm pc}}
\def \msol{\mathrm{M}_{\odot}}

\def \yr {\mathrm{yr}}

\def \K {\mathrm{K}}
\def \mvir {M_{\mathrm{vir}}}

\def \nhi {N_{\mathrm{HI}}}
\newcommand{\dex}{\mathrm{dex}}

\definecolor{DarkGreen}{rgb}{0.5,0.,0}

\def\lsim{\mathrel{\rlap{\lower3pt\hbox{$\sim$}}
    \raise1pt\hbox{$<$}}}                % less than or approx. symbol
\def\gsim{\mathrel{\rlap{\lower3pt\hbox{$\sim$}}
    \raise1pt\hbox{$>$}}}                % greater than or approx. symbol

\begin{document}

\pubyear{2009}
\title[Simulating GRB afterglow absorbers]{The nature of \hi\, absorbers in GRB afterglows: \\ clues from hydrodynamic simulations}
\author[A. Pontzen et al.]{Andrew Pontzen$^{1}$\thanks{Email:
    apontzen@ast.cam.ac.uk}, Alis Deason$^{1}$, Fabio Governato$^{2}$, Max Pettini$^{1}$,
James Wadsley$^{3}$,  \newauthor Thomas Quinn$^{2}$, Alyson Brooks$^{2,4}$, Jillian Bellovary$^{2}$, %\newauthor
Johan P. U. Fynbo$^{5}$ \\
%C.M. Booth$^{3,4}$  \newauthor Greg Stinson$^{2,5}$,
%James Wadsley$^{5}$,  Alyson Brooks$^{2}$, Thomas Quinn$^{2}$ \\
  $^{1}$Institute of Astronomy, Madingley Road, Cambridge CB3 0HA, UK \\
  $^{2}$Astronomy Department, Box 351580, University of Washington,
  Seattle, WA 98195, USA \\
  $^{3}$Department of Physics and Astronomy, McMaster University, Hamilton, ON L8S 4M1, Canada \\
  $^{4}$California Institute of Technology, M/C 130-33, Pasadena, CA 91125, USA \\
  $^{5}$Dark Cosmology Centre, Niels Bohr Institute, University of Copenhagen, Juliane Maries Vej 30, DK-2100 Copenhagen, Denmark}

\date{Accepted 2009 November 10. Received 2009 November 10; in original form 2009 August 25.}
\maketitle

\begin{abstract}
  In recent work, we have shown that it is possible to link
  quantitatively many aspects of damped Lyman alpha (DLA) absorbers in
  the spectra of quasars to high resolution simulations of galaxy
  formation.  Using runs from the same series of hydrodynamic
  numerical studies, we consider the expected properties of intrinsic
  Lyman alpha absorbers seen in the spectra of high redshift ($z>2$)
  gamma ray burst afterglows (GRB-DLAs). If GRBs are associated with
  the death of massive stars, their afterglows provide insights into
  otherwise unprobed regions of protogalactic objects, but detailed
  physical interpretations are currently embryonic.
  
We find that median impact parameters (measured from the potential
minimum) are approximately $1 \kpc$ for GRBs compared with $4 \kpc$
for QSO-DLAs. However, an equally important difference is that
GRB-DLAs are predominantly associated with halos of mass
$10^{10}<M_{\mathrm{vir}}/M_{\odot}<10^{12}$, an order of magnitude
larger than the hosts of QSO-DLAs. Accordingly, there are differences
in the stellar properties of hosts. For instance mean star formation
rates are higher: $\langle \dot{M}_{\star} \rangle \simeq 10 \, \msol
\yr^{-1}$ for GRB-DLAs compared with $\langle \dot{M}_{\star} \rangle
\simeq 1 \, \msol \yr^{-1}$ for QSO-DLAs.

  Our simulations accurately predict the form of the GRB-DLA
  \hi~column density distribution, producing quantitative agreement
  for $\nhi>10^{19} \cm^{-2}$, but they somewhat underpredict the
  incidence of low column densities $N_{\mathrm{HI}}<10^{19}
  \cm^{-2}$. This is reflected in our estimate of the ionizing photon
  escape fraction, $f_{\mathrm{esc}} \simeq 1\%$, which is lower than
  the observational GRB-derived escape fraction ($2\%$).
  Line-of-sight neutral gas metallicities predicted by our simulations
  ($10^{-2}<Z/Z_{\odot}<1$) are consistent with the modest
  observational constraints. Because of large internal dispersions in
  gas metallicities, this agreement is not significantly compromised
  by imposing a cut-off on the metallicity of stars able to launch
  GRBs ($Z_{\star}<Z_{\odot}/3$), confounding claims that the observed
  metallicity of GRB-DLAs poses a severe challenge to current GRB
  models.

\end{abstract}

\begin{keywords}
gamma-rays: bursts -- quasars: absorption lines -- galaxies: formation
\end{keywords}

\section{Introduction}\label{sec:introduction}

\begin{figure}\label{fig:cartoon}
\includegraphics[width=0.5\textwidth]{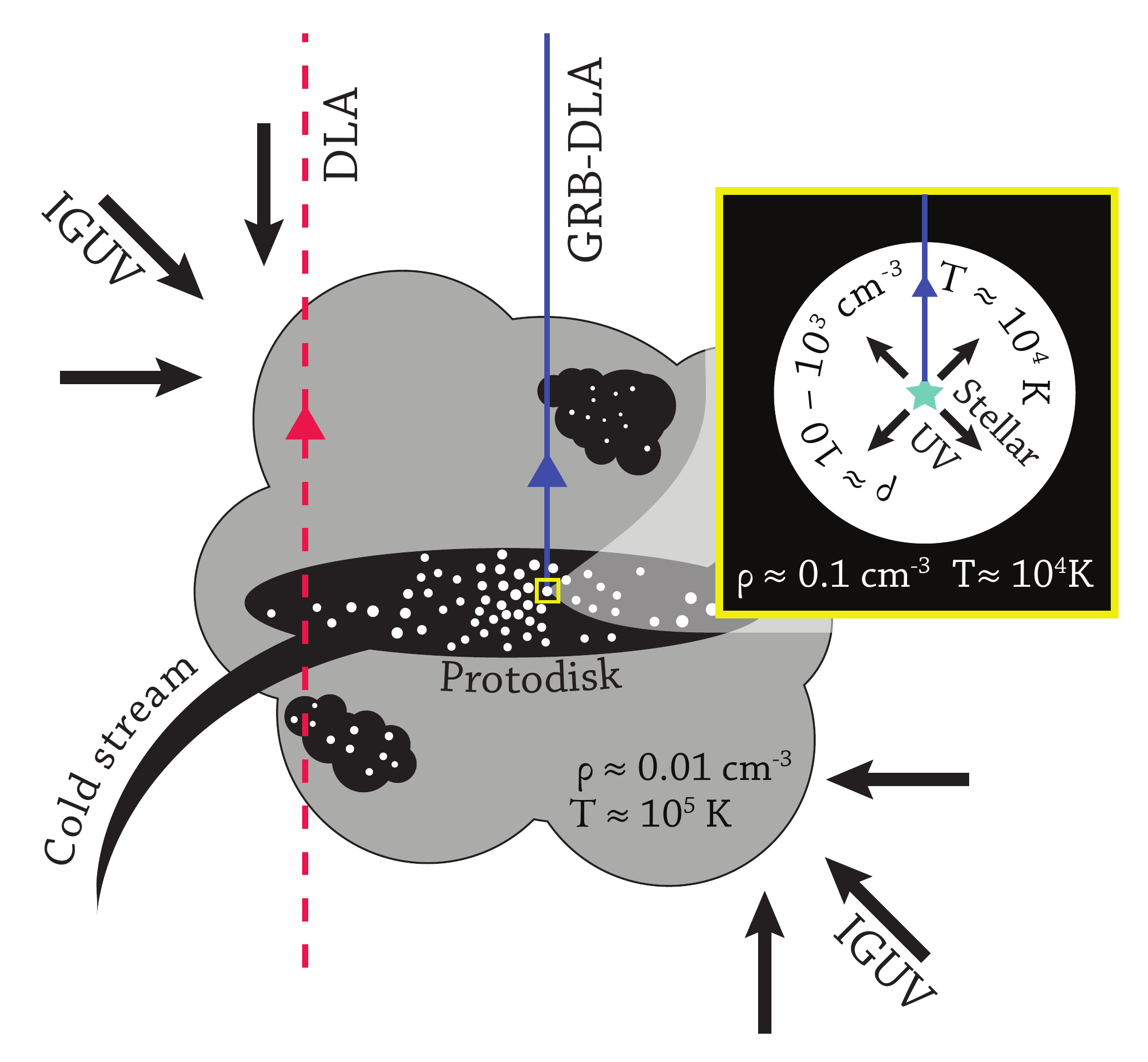}
\caption{A sketch of relevant differences between typical QSO (dashed
  vertical line) and GRB afterglow (solid vertical line) sightlines
  through a protogalaxy at high redshifts ($z \sim 3$), after
  \protect\cite{2007ApJ...666..267P}. Assuming GRBs are associated
  with the death of massive young stars, this class of object arises
  inside dense \hii~regions (white circles) within the the most
  rapidly star forming regions of the neutral medium (solid black
  regions); whereas QSO sightlines are more likely to intersect the
  neutral medium at a larger impact parameter. In our simulations,
  single protogalactic halos (particularly those with
  $M_{\mathrm{vir}}\gsim 10^{10.5}\, \msol$) can host multiple neutral
  star-forming regions; thus GRB sightlines may also intersect a
  separate neutral zone which has little physical association with
  their origin. The local stellar UV, presumed largely absorbed within
  the \hii~regions by our model (Section \ref{sec:radi-trans}), has an
  effect on GRB-DLAs but appears unimportant in understanding the
  QSO-DLAs. However the incoming background intergalactic mean field
  has an impact on both classes of object (Section
  \ref{sec:radi-trans}).  Typical temperatures and densities for the
  regions of interest are indicated (based on observations of our
  own Galaxy).}
\end{figure}

Observations of transient Gamma Ray Burst (GRB) afterglows have
recently been established as a promising new window on the high
redshift Universe. In particular the coupling of rapid response
observing opportunities on large telescope facilities with the
detection alerts provided by the Swift satellite
\citep{2004ApJ...611.1005G} have produced a modest but growing sample
of high quality data.  The GRBs are assumed (in the case of long
duration GRBs which have brighter optical afterglows and will be the
focus of the present work) to be associated with the death of
Wolf-Rayet stars \citep{1999ApJ...524..262M}.

Intrinsic absorption features in the GRB afterglow spectra are of
particular interest because they probe the gaseous environment of the
progenitor. The physical state of the gas may be quite different from
that of typical (cross-section weighted) intervening absorbers. Dense
neutral hydrogen giving rise to intervening absorption with column
densities $\nhi$ exceeding $2 \times 10^{20} \, \cm^{-2}$ is known as
a damped Lyman alpha system (DLA); the \hi~absorption and its
associated metal line features are commonly studied in the spectra of
quasars \cite[e.g.][]{2005ARA&A..43..861W}. The focus of the present
paper will be the analogue of DLAs seen in the intrinsic spectra of
GRB afterglows: neutral hydrogen absorption at the emission redshift
of the GRB itself, by analogy termed a ``GRB-DLA''\footnote{For
  simplicity this term is applied to all intrinsic GRB absorption
  systems even though $20\%$ would not exceed the traditional DLA
  threshold.}.  Because GRB-DLAs are physically associated with the
GRB hosts, their nature may be quite different from intervening DLAs
studied in quasar spectra (QSO-DLAs); part of the aim of this paper is
to recap known observational differences and link them to the
underlying protogalactic populations.

Indeed, a statistical comparison of QSO-DLAs with GRB-DLAs immediately
confirms that the populations are distinct. While the frequency of
quasar absorption systems decays monotonically with increasing neutral
hydrogen column density
\cite[e.g.][]{2005ARA&A..43..861W,2005ApJ...635..123P}, the observed
frequency of GRB-DLAs grows to a peak at $N_{\mathrm{HI}} \simeq
10^{22} \cm^{-2}$ (a column density exceeding the strongest QSO-DLA
ever observed; \citeauthor{2006A&A...460L..13J} 2006). Metallicities
of the \hi-associated gas, although challenging to measure in the
transient GRB-DLA afterglows, are generally in the range $1/100 <
Z/Z_{\odot} < 1$, with a median somewhat exceeding $Z_{\odot}/10$
\cite[e.g.][]{2006NJPh....8..195S,2007ApJ...666..267P,2006A&A...451L..47F};
whereas the median QSO-DLA metallicity is around $Z_{\odot}/30$
\citep{2006fdg..conf..319P}.

The physical interpretation of the GRB-DLA population is maturing, but
is not firmly established. It seems clear that, qualitatively, an
association of GRB events with the death of massive, short-lived stars
leads to the correct expectation of higher column densities (to ensure
high star formation rates) and higher metallicities.  This has led to
the schematic picture \cite[e.g.][]{2007ApJ...666..267P} that GRB-DLAs
preferentially probe the innermost regions of rapidly star forming
protogalactic halos (Figure \ref{fig:cartoon}), while QSO-DLAs
intersect the outer regions of such systems.  The neutral gas probed
by the GRB-DLAs is not thought to be in the immediate circumstellar
environment but rather at distances exceeding at least 10
parsecs. This follows because, under realistic assumptions, the UV
emission from the afterglow itself is capable of ionising any local
neutral material within just an hour of the photons from the initial
burst arriving \citep{2006ApJ...648...95P}. The non-local picture is
reaffirmed by the non-detection of time variations in the neutral
columns \citep{1998ApJ...501..467P,2002ApJ...578..818M}. In addition,
approximate upper limits on distances of absorbing gas can be derived
from UV pumping of fine-structure transitions, which suggest the
responsible neutral gas is at a distance of at least $100\,\pc$
\citep{2006ApJ...648...95P} and perhaps as much as $\sim 1\,\kpc$
\citep{2007A&A...468...83V,2009ApJ...694..332D}. Finally, dust
depletion patterns seem more consistent with those seen in the warm
neutral ISM of our Galaxy, rather than in cold dense molecular clouds
\citep{2006NJPh....8..195S}.

It is possible that, in addition to the GRB-DLAs probing particular
regions of halos, the population of objects giving rise to GRB-DLAs
are on average more massive than those giving rise to QSO-DLAs. This
will arise if the cold gas cross-section (relevant for QSO-DLAs) has a
shallower than linear relationship with the UV luminosity (assumed
relevant for GRB-DLAs); such a picture has been successfully employed
by \cite{2008ApJ...683..321F} to reproduce differences in
metallicities of the absorber populations.

As well as providing insights into individual objects, GRB-DLAs offer
an unusually direct way to probe the escape of ionizing photons from
the star forming regions with which they are associated. On the
assumption that the GRB rate traces the star formation accurately, one
may construct a global escape fraction $\langle f_{\mathrm{esc}}
\rangle \simeq 2\%$ by modest extrapolation from the GRB-DLA column
density distribution
\citep{2007ApJ...667L.125C,2009arXiv0907.3449F}. Conversely recent
simulations by \cite{2008ApJ...672..765G}, aimed at constraining
escape fraction statistics, have incidentally been able to account for
the GRB-DLA distribution (albeit with some mismatch at high column
densities).

The emphasis in our work is reversed: we focus on statistics of
GRB-DLAs between $2<z<4$, $90\%$ of which have column densities
exceeding $10^{19} \cm^{-2}$.  We use high resolution hydrodynamical
simulations with physically motivated star formation feedback
algorithms \citep{2006astro.ph..2350S} to generate a catalogue of
GRB-DLA sightlines.  This catalogue can be compared to current
observational constraints, while the association of each simulated
sightline with a specific protogalactic object allows us to predict
the relationship between GRB-DLAs and other known classes of objects
\cite[see also][]{2008ApJ...683..321F}. We can regard such
associations with some confidence since, after calibrating two free
parameters in isolated disk galaxy tests, the simulations are known to
reproduce realistic $z=0$ disk galaxies
\citep{2007MNRAS.374.1479G,2008arXiv0812.0379G}, have realistic
stellar mass-metallicity relations at low and high redshifts
(\citeauthor{2007ApJ...655L..17B} 2007; see also section 7.5 of
\citeauthor{2008A&A...488..463M} 2008), and have produced a population
of QSO-DLAs which are as realistic as current techniques permit
\citep[for a detailed comparison to observations see][henceforth
  P08]{PontzenDLA}. The treatment of local ionising sources we adopt
is simple compared to detailed modelling employed in some ionising
photon escape fraction studies
\cite[e.g.][]{2006ApJ...651L..89R,2008ApJ...672..765G,2009ApJ...693..984W}.
As a result, the statistics of systems with $\nhi<10^{19} \cm^{-2}$
seem to be sensitive to assumptions in our approximations, and
corresponding caution will be exercised in our interpretation -- but
results appear robust for larger column densities.

The simulations employed and techniques applied in the present work
are closely related to those of P08, and are discussed further in
Section \ref{sec:techniques}.  By tracing sightlines from star forming
regions we produce a set of simulated GRB-DLA statistics which we
compare, where possible, with existing observations (Section
\ref{sec:results}). While both observational statistics and
theoretical models are far less certain for GRB-DLAs than QSO-DLAs, we
are able to draw tentative conclusions from our present study in
Section \ref{sec:conc-dis}.
 
\section{Simulations and Techniques}\label{sec:techniques}

In this Section, we describe the simulations employed and discuss the
techniques required to build a cosmological sample of GRB-DLAs.

The simulations considered here are similar to those employed by
P08. In particular, they take advantage of the `volume
renormalization' technique, in which the dynamic range is expanded by
computing a subregion of the full volume at high resolution
\citep{1993ApJ...412..455K}: all results are taken from the high
resolution zone. Time evolution is computed with the SPH code
\textsc{Gasoline} \citep{2004NewA....9..137W} which implements gas
cooling, the effects of a cosmological UV background
\citep{1996ApJ...461...20H} with a local approximate self-shielding
algorithm (see section 5.4 of P08), and the effects of star formation
based on the description of \cite{2006astro.ph..2350S}. Supernova
ejecta abundances for oxygen and iron are calculated in the simulation
following the yields of \cite{1986A&A...158...17T} and
\cite{1993PhR...227...65W}; after being injected into the interstellar
gas these metals are advected with the SPH particles, although we now
incorporate a correction for sub-resolution turbulent mixing which can
transfer metals between Lagrangian fluid elements
\cite[see][]{2009arXiv0910.5956S}. Gravitational softening is constant in
physical units for $z<8$ and evolves comovingly for $z>8$; its final
fixed value is given for each simulation in Table
\ref{tab:simulations}. For a full description see section 2 of P08.

The particular regions in use for the present work are listed in Table
\ref{tab:simulations}. These mirror closely the regions used by P08,
although none are in fact identical. The first three are selected to
form (by $z=0$) individual objects and their immediate environments:
`DG' is a high resolution run which, at $z=0$, forms a dwarf galaxy ($\mvir
\simeq 3 \times 10^{10} \,\msol$) at $z=0$; `MG' forms a Milky Way-like
galaxy ($\mvir \simeq 7 \times 10^{11}\, \msol$) and `CL'
forms a cluster ($\mvir \simeq 1.1 \times 10^{13} \msol$). `C50' is a full cosmological box of side $50\,\Mpc$ (comoving).
From each box we select, at $z=3$, the set of halos having greater
than $3\,500$ dark matter particles -- an empirical limit which we
established ensured convergence of our results including stellar
populations \citep[see also P08;][]{2007ApJ...655L..17B}. The `CL'
simulation, combining high resolution with large volume, contains 86
halos satisfying this stringent criterion whereas the three other
simulations have only $\sim 15$ each. For this reason, we selected a
random subsample of $15$ halos from the `CL' simulation for our
statistical results which are therefore derived from $60$ halos of
high resolution, evenly distributed with
$10^8<M_{\mathrm{vir}}/\msol<10^{12.5}$. We discuss the impacts of the
finite mass range of resolved halos in Section
\ref{sec:origins-grb-dlas}.

As always with `zoomed' simulations, one must be as careful as
possible to test whether the arbitrary choice of region may have
affected results through environmental dependences; we take agreement
of statistics between boxes (where their resolved halo mass ranges
overlap) as evidence that no severe problems are to be expected. In
particular our present simulations include at $z=3$ an over-dense
proto-cluster region (`CL') which resolves the same range of halo
masses as the more typical `MG' box. There is no evidence for
meaningful disagreement between the properties of halos within these
boxes \cite[see also][]{2009arXiv0906.4350C}. However, as in P08, we
must accept in principle the possibility that we miss an environmental
effect which has an impact on statistics -- this is to be regarded as
a trade-off against resolving such a large range of halo masses with
relatively uniform resolution.

While the regions are from a different realization of the initial
power spectrum than those presented in P08, the physics remains almost
identical except for two notable differences. Firstly we have used the
self-shielding approximation, described in section 5.4 of P08, in all
our simulations. This has the effect of slightly raising cold gas
fractions and lowering dwarf galaxy metallicities (the latter because
the lower pressure of the ambient ISM makes our feedback algorithm
somewhat more effective); see P08 for details. Secondly, we have
implemented a turbulent metal diffusion algorithm as described at the
start of this section.

Because of the long lead-time involved in running the simulations
considered, the cosmological parameters employed
($\Omega_{\mathrm{M}}=0.24$, $\Omega_{\Lambda}=0.76$,
$\Omega_b=0.042$, $h=H_0/(100\,\kms)=0.73$, $\sigma_8=0.77$, $n_s =
0.95$) are close to, but not fully, consistent with the latest
constraints from large scale observational results
\cite[e.g.][]{2008arXiv0803.0586D}. As we have previously argued, we
do not believe small differences in the adopted cosmology will impact
strongly on the results (see section 5.3 of P08).

\begin{table}
\begin{center}
\begin{tabular}{lllll}
\hline
Tag & $\max M_{\mathrm{p,gas}}$ & $ M_{\mathrm{p,DM}} $ & $\epsilon/\mathrm{kpc}$ &  Usable Vol (comoving) \\
\hline
DG & $10^{3.9} \mathrm{M}_{\odot}$ & $10^{4.6} \mathrm{M}_{\odot}$ & $0.12$ & $4\, \Mpc^3$ \\ 
MG & $10^{5.6}$ & $10^{6.0}$ & $0.34$ & $70\, \Mpc^3$ \\
CL & $10^{5.6}$ & $10^{6.0}$ & $0.34$ &  $600\, \Mpc^3 $ \\
C50 & $10^{7.9}$ & $10^{8.4}$ & $2.00$ &  $125\,000 \Mpc^3$  \\
\hline
\end{tabular}
\end{center}
\caption{The simulations used in this work. The first column is the
  tag which we use to refer to each simulation. For all except
  ``C50'', a subsample of the full box is simulated in high
  resolution; no results are taken from outside this region.  The
  second and third columns refer respectively to the maximum gas and dark
  matter particle masses within the region, the fourth to the
  gravitational softening length (in physical units) and the final
  column gives the comoving volume of the region. The separate boxes
  are generated from entirely different sets of initial conditions;
  the ``DG'' and ``MG'' simulations are designed to form respectively
  a dwarf and approximately $L_{\star}$ galaxy at $z=0$ while the
  ``CL'' region forms a large cluster of galaxies at $z=0$.
}\label{tab:simulations}
\end{table}

\subsection{Approximate Radiative Transfer}\label{sec:radi-trans}

Understanding the effect of radiation on the ionization state of the
gas is crucial to obtaining realistic absorption statistics. The
simulations incorporate a diffuse cosmological UV background based on
\cite{1996ApJ...461...20H}, uniform except in dense regions for which
a local self-shielding algorithm, described in Section 5.4 of P08, is
applied. We post-process the output with a simple equilibrium
radiative transfer algorithm, described in Section 3.1 of P08, to
improve our estimate of the self-shielding effect. We previously
argued that the correction to this picture from local sources, which
were not included, would be small.

However, in GRB-DLAs, where the sightline necessarily passes through
highly ionized bubbles around hot stars, local radiation sources are
likely to have a stronger impact on results (these hot bubbles
presumably having negligible cross-section to a QSO
sightline). Observationally, fine-structure transitions in GRB-DLA
absorbers indeed suggest the presence of a local UV field
\citep{2007ApJ...666..267P}.

A study of the local ISM shows that incorporating the stellar ionizing
sources into a radiative transfer scheme using the raw density field
from the simulations will overestimate the effect of local
sources. This is because young stars are initially surrounded by a
dense molecular cloud, which is rapidly photoevaporated to form an
\hii~region. Such regions have high densities, typically between $10$
and $100\,\cm^{-3}$ but sometimes as high as $10^4\,\cm^{-3}$.  Hence
their recombination time is short and they are able to absorb the vast
majority of the UV radiation produced within them. This micro-physics
occurs on parsec scales, below the resolution limit of our
simulations.

Assuming typical temperatures and densities for \hii~regions
($T=10^4\,\K$, $n_{\mathrm{H}} = 30 \cm^{-3}$), one may estimate the
hydrogen mass required to fully absorb all UV photons produced, using
an on-the-spot approximation to account for photons emitted by
recombination to the ground state. Using \textsc{Starburst99}
\citep{2005ApJ...621..695V} emission models we find

\begin{equation}
M_{\mathrm{HI\to HII}}\simeq 1.5 \times 10^7 \msol \left( \frac{\dot{M}_{\star}}{1 \msol \yr^{-1}}\right) \left( \frac{30\,\cm^{-3}}{n_{\mathrm{H}}}\right)\label{eq:sf-uv}
\end{equation}
of neutral hydrogen would be ionized by the local sources, where
$\dot{M}_{\star}$ is the star formation rate, the final factor is
fixed by assumption in our calculations and a
\cite{1993MNRAS.262..545K} initial mass function is assumed for
consistency with the simulations.

To assess the impact of removing the corresponding neutral material,
we calculate the total star formation rate per halo and expand ionized
spheres around all young stellar particles ($<10^7 \yr$) until the
total mass converted from \hi~to \hii~is given by
(\ref{eq:sf-uv}). Because the star forming threshold density ($0.1\,
\cm^{-3}$) in the simulations are lower than true \hii~region
densities, the volume affected by this algorithm is rather larger than
the physical sizes of \hii~regions would suggest, but by construction
the mass affected is accurate. This should therefore reduce
line-of-sight column densities by an appropriate
fraction\footnote{There is some evidence to suggest that the GRBs
  themselves ionise material along their line-of-sight (see
  Introduction); with our current knowledge this is hard to model in
  detail, but we discuss possible consequences for our results in
  Section \ref{sec:conc-dis}).}.  We verified that the impact on the
quasar DLA statistics is, as expected, insignificant, but the effect
on GRB-DLAs is non-negligible (Section \ref{sec:results}).

%[Sub-resolution (hiis) method]
%
%We compute the star formation rate for each gas particle
%  according to the simulation's criteria ($\dot{M}_{\star,p} =
%  c_{\star} M_{\mathrm{gas}} \sqrt{G \rho_{\mathrm{gas}}}$ with
%  $c_{\star}=0.05$ in converging flows with $T<3\times 10^4 K$,
%  $n_{\mathrm{gas}}>0.1 \cm^{-3}$) and reduce the neutral fraction for
%  that particle such that the amount of \hi~lost is given by
%  (\ref{eq:sf-uv}). (This fraction is always less than $100\%$ for the
%  relevent densities.) This implicitly places the \hii~ in clumps
%  below the resolution limit, leaving a sparser \hi~warm neutral
%  medium.  This method has the advantage that it could be suitable for
%  implementation in a live simulation. However it would not naturally
%  yield the correct limiting behaviour (of resolved \hii~regions
%  around individual stars) if the resolution is increased. Further,
%  under this scheme a given particle no longer has a straight-forward
%  interpretation as homogeneous gas in a fixed state.

%We found that the two techniques gave results in fair agreement
%(although the effect of {\it i} is slightly stronger than that of {\it
%  ii}, since it concentrates the same mass conversion in a smaller
%region), and display only results using method ({\it i}); we also

Our approach ignores the effect of dust which is capable of absorbing
a significant fraction of ionizing radiation
\citep{2006agna.book.....O}; since the quantity of dust in high
redshift \hii~regions is hard to assess, we merely admit that the
magnitude of our correction could be reduced by including dust (see
also Section \ref{sec:conc-dis}). We further fail to account for
molecular clouds, which could be important in understanding DLAs and
GRB-DLAs \citep[e.g.][]{2001ApJ...562L..95S}. The small fraction of
detectable H$_2$ in most QSO-DLAs \citep{2003MNRAS.346..209L} and lack
of H$_2$ in many GRB-DLAs (\citeauthor{2007ApJ...668..667T} 2007;
although see \citeauthor{2006A&A...451L..47F} 2006;
\citeauthor{2009ApJ...691L..27P} 2009) does not imply H$_2$ is
unimportant, as large masses could conceivably reside in small clouds
with negligible cross-section. \cite{2009arXiv0907.1057L} also
suggested that non-detections in most GRB spectra are expected given
the low metallicities of typical systems observed with high resolution
spectroscopy. Overall the fraction of gas in H$_2$ clouds is, at
present, not calculable from our simulations and so we are forced to
proceed on the assumption that later inclusion will make only a small
correction to our results. Further discussion of the inherent problems
in modelling the small-scale ISM and its impact on our results is
given in Section \ref{sec:conc-dis}.

\subsection{Sightline Generation}\label{sec:sightline-generation}

The theoretical and observational picture of the precise origins of
long GRBs is complex \cite[e.g.][and references
  therein]{2006ARA&A..44..507W}. We will suppose the events to be
associated with the death of massive stars ($\sim 25 \msol$); such
stars have a main sequence lifetime of just a few million years, far
smaller than the dynamical time of host halos (e.g. for the adopted
cosmology at $z=3$, $(200\,G\rho_c)^{-1/2} \simeq 7 \times 10^8\,
\yr$).  For the purposes of this work, long GRBs are therefore assumed
to arise in a given region with probability proportional to the star
formation rate of that region. We choose the star particles younger
than $5 \times 10^{7} \yr$ in each halo as launch sites for our tracer
sightlines in order to reflect this assumed association. Clearly the
maximum age exceeds by an order of magnitude the actual lifetime of
the GRB progenitors, but it is required to guarantee at least $200$
candidate launch sites in each star forming halo (in turn required for
reliability of our resampling mechanism described in Section
\ref{sec:resampling}).  We verified that reducing cut-off times to
$10^{7} \yr$ (the minimum possible at our resolutions) did not
significantly affect our results with the exception of the low column
density distribution ($<10^{19}\,\cm^{-2}$) -- see
Section~\ref{sec:column-dens-distr}.

% %%%%%%%%%% Method 2
%
%Operationally this is achieved by
%calculating the star formation rate by mass for each gas particle (see
%above), then picking sightline origins in proportion to that rate.  To
%allow stars to drift relative to their parent gas clouds, we added a
%gaussian random offset with standard deviation $\Delta = 100\, \pc$ to
%each chosen progenitor origin. This corresponds, for
%$\tau_{\mathrm{MS}} = 6\, \Myr$, to a velocity dispersion
%$\sigma\simeq 15\,\kms$ relative to the originating gas cloud. We
%verified that our results are insensitive to varying $\Delta$ within
%the range $50<\Delta/\pc<200$.

Additionally, we considered the effect of a metallicity ceiling for
progenitor stars; such a limit is motivated by GRB collapsar models
which require mass loss over the progenitor Wolf Rayet stellar
lifetime to be small \citep{1999ApJ...524..262M}. However, we found
its effect -- even on the measured column metallicities -- to be
minor.  A discussion is given in
Section~\ref{sec:metal-ceiling}.

The sightline properties are generated using the code described in
P08, performing a full SPH line-of-sight integration to determine
the column densities of neutral hydrogen and neutral-phase metals, i.e.
the sightline metallicity is defined to be 
\begin{equation}
Z_{\mathrm{sightline}} = \frac{\int_0^{\infty}  \dd x\, Z(x)\,n_{\mathrm{HI}}(x)}{\int_0^{\infty} n_{\mathrm{HI}}(x)}\label{eq:sl-metal}
\end{equation}
where $x$ is the distance from the GRB along the line of sight and
$Z(x)$ is the metal mass fraction at that point.  Equation
(\ref{eq:sl-metal}) makes the implicit assumption that observers
measure metal ions which are perfectly coupled\footnote{This assumption was also made
  for generating \siii~profiles in P08; however contrary to the
  statement in \citeauthor{2009arXiv0904.3545T} 2009, we did not
  assume a fixed metal {\it abundance} when generating
  \siii~profiles. We also used \textsc{Cloudy} models to show that the
  correction to the perfect coupling model is small (P08 section
  5.2).} to the \hi~(e.g., for
silicon, that $n_{\mathrm{Si\,II}}/n_{\mathrm{Si}} =
n_{\mathrm{HI}}/n_{\mathrm{H}}$).

We use the oxygen mass abundance reported by the simulations as our
metallicity tracer, using the reference value
%$(M_{\mathrm{O}}/M_{\mathrm{H}})_{\odot} = 0.0078$
$12+\log_{10}(n_{\mathrm{O}}/n_{\mathrm{H}})_{\odot} = 8.69$
\citep[from][]{2003ApJ...591.1220L} to normalize to a solar
metallicity scale. The choice of oxygen reflects its status as the
most abundant metal by mass; differences between elements are small
relative to the overall spread of metallicities under consideration.

\subsection{Resampling}\label{sec:resampling}

The set of halos extracted from all boxes is used to generate a
cosmological sample in a directly analogous fashion to that described
by P08. However, instead of using a gaseous cross-section we arrange
that the probability of selecting a sightline from a particular halo
of virial mass $M_h$ is proportional to the product of the halo mass
function $f(M_h)$ and the star formation rate of that halo
$\dot{M}_{\star,h}$.  We bin our ensemble of halos in virial mass; the
probability of a given GRB originating in the halo $h$ from mass bin
$i(h)$ is then proportional to $w_h$,
\begin{equation}
w_h = \frac{F_{i(h)} \dot{M}_{\star,h}}{n_{i(h)}}\textrm{.} \label{eq:weight}
\end{equation}
where $F_i$ is the halo mass function integrated over the mass range
of bin $i$ (P08 equation 4) and $n_i$ is the total number of halos
from our simulations in that bin. All equations from P08 follow
identically with our replacement definition (\ref{eq:weight}) for the
halo weight $w_h$.

\section{Results}\label{sec:results}

\subsection{Origins of GRB-DLAs}\label{sec:origins-grb-dlas}

\begin{figure}
\includegraphics[width=0.5\textwidth]{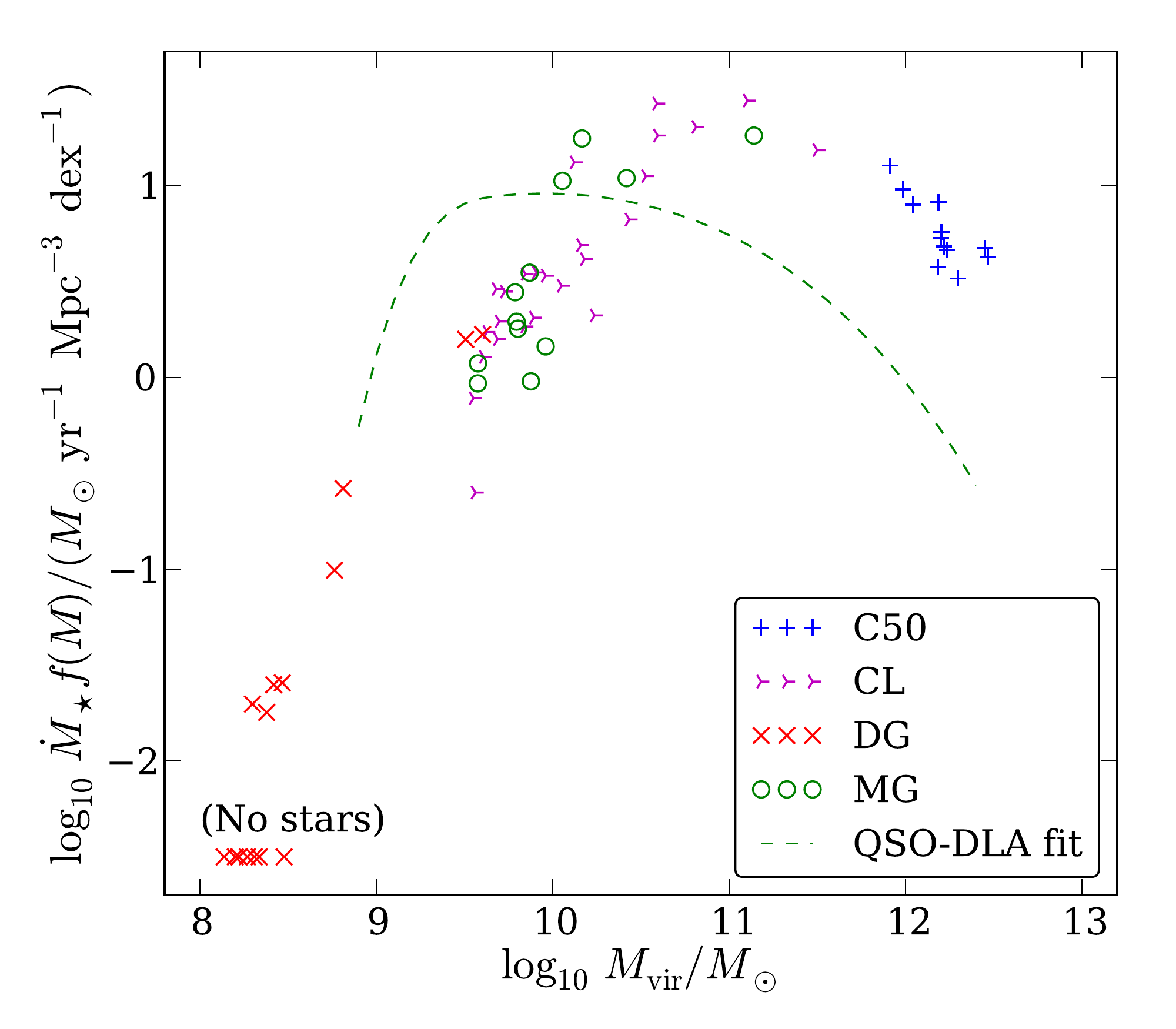}
\caption{The product of the star formation rate and halo mass function
  for each qualifying halo (see Section \ref{sec:techniques} for
  resolution criteria) in our DG (crosses), MG (circles), CL (tripods)
  and C50 (plus symbols) runs.  This quantity is proportional to the
  contribution of a given halo mass to the overall cosmological
  ensemble of GRBs and therefore GRB-DLAs. A fit to the equivalent
  quantity for quasar DLAs is shown as a dashed line, based on the
  locus of points in figure 5 of P08. Halos with no star formation are
  grouped together at the bottom of the figure.}\label{fig:xsec}
\end{figure}

Because GRB-DLAs are intrinsic to the galaxy hosting the GRB event
itself, the cosmological regions probed by such absorbers are not
necessarily the same as those probed by the population of intervening
QSO-DLAs.  By assumption, our method triggers GRBs in a given region
proportionally to that region's contribution to the cosmological star
formation rate. This can be split into two factors: the relative
contribution of halos of different masses, and the position of the
star forming regions within each halo. In this section we will
consider each of these two contributing effects in turn.

Figure \ref{fig:xsec} shows the product of the halo mass and star
formation rate ($f(M_h)\,\dot{M}_{\star,h}$) for individual halos in
each of our boxes; as described in Section \ref{sec:resampling}, this
relationship defines the relative contribution of halos of the
specified virial mass. We find that a broad range of halos contribute
to the overall population of GRB hosts, peaking at $\sim 10^{11}
\msol$. The distribution is in good agreement between our four
separate boxes in their regions of overlap. In particular the
proto-cluster (CL) environment shows no systematic differences from
the disk galaxy progenitor region (MG).

For masses $\mvir<10^{9.5} \msol$, the contribution tails off as star
formation rates are stifled by the stellar feedback and UV field
\citep[e.g.][P08]{2006MNRAS.371..885R}. The steep drop at low masses
in Figure \ref{fig:xsec} shows that our low-mass cutoff of $\mvir=10^8
\,\msol$ should not lead to any systematic effects. For
$\mvir>10^{11.5}\,\msol$ the GRB rate contribution begins to be
suppressed by the downturn in the cosmological halo mass function;
however, somewhat unsatisfactorily, our most massive halo has $\mvir =
10^{12.5} \msol$ which means that the volume of our simulations is
insufficient to trace objects until they are so rare as to be
cosmologically insignificant. We checked for systematic effects by
artificially cutting off our distribution at $10^{12} \msol$, finding
that no results were severely effected, but a satisfactory resolution
would require considerably larger simulations to produce a full
population of rare $\mvir>10^{12.5} \msol$ objects.

We have overplotted (dotted line in Figure \ref{fig:xsec}) a fit to
the equivalent halo-weighting quantity for QSO-DLAs,
$\sigma_{\mathrm{DLA}} f(M)$ where $\sigma_{\mathrm{DLA}}$ is the
cross-section of the halo for column densities of neutral hydrogen
$\nhi>10^{20.3} \cm^{-2}$ from P08. The dominant contribution to the
QSO-DLAs can be seen to arise in halos approximately ten times less
massive than for GRB-DLAs; the mechanisms determining the upper and
lower limits are similar in both cases, but because
$\sigma_{\mathrm{DLA}}$ scales less steeply than $\dot{M}_{\star}$
($\sigma_{\mathrm{DLA}} \sim M_{\mathrm{vir}}$, P08 figure 4;
$\dot{M}_{\star} \sim \mvir^{1.6}$, P08 figure 13), the latter
distribution is skewed towards higher masses. We will comment in
section \ref{sec:metall-distr} that this has important consequences
for the metallicity distribution of our simulated GRB-DLAs.

\begin{figure}
\includegraphics[width=0.5\textwidth]{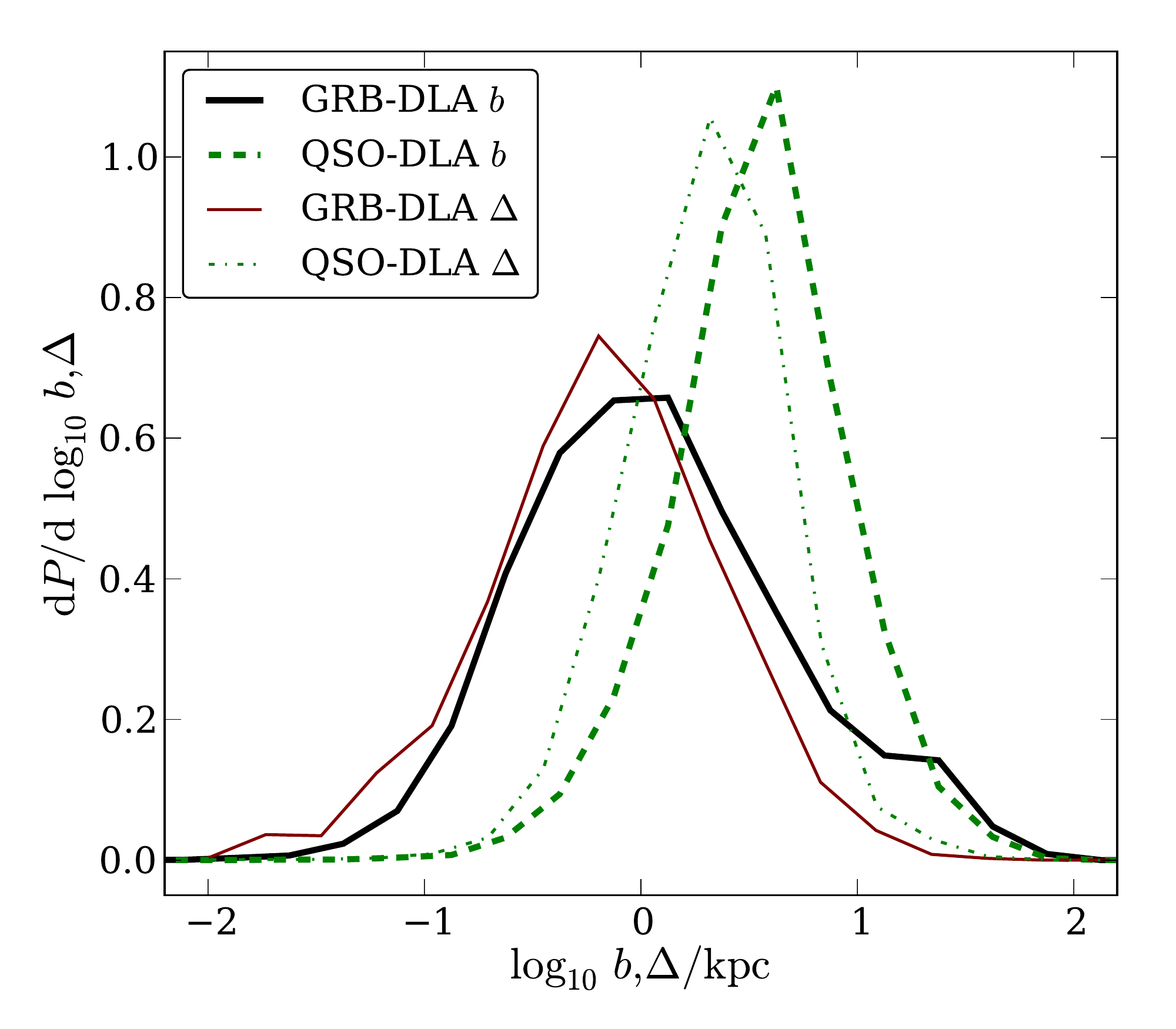}
\caption{The distribution of impact parameters $b$ for GRB-DLAs (solid
  thick line) and QSO-DLAs (dotted thick line). The GRB-DLA population
  typically originate very close to the centre of the dark matter
  halo, whereas the QSO-DLAs tend to pass several kpc from the centre,
  although in both cases there is a long tail to large impact
  parameters. This tail is associated with massive
  ($\mvir>10^{11}\,\msol$) halos with recent merging and hence
  multiple gas/star formation centres. The longitudinal distances,
  $\Delta$, over which the absorbers are spread (see text for exact
  definition) are also shown as thinner lines; the GRB-DLAs are seen
  to be more compact along the line-of-sight. All units are
  physical.}\label{fig:impact}
\end{figure}

We now turn to the location of GRB-DLAs within individual
halos. Figure \ref{fig:impact} shows the impact parameter $b$
(projected distance of the DLA sightline to the minimum gravitational
potential of the host halo) for GRB-DLAs (thick solid line) and
QSO-DLAs (thick dashed line) in our current set of simulations. The
former arise considerably closer to the halo centre (typically at
$\sim 1 \kpc$) than the latter (at $\sim 4 \kpc$; although there is a
spread over orders of magnitude in distance for both cases). The most
active star formation is naturally associated with the deepest part of
the potential well, so that these results can be interpreted as
confirming the intuitive picture of Figure \ref{fig:cartoon}.

In both the QSO and GRB case there is a tail of impact parameters
extending to $100 \kpc$ distances, arising in high mass halos
($M_{\mathrm{vir}}>10^{11} \msol$) which are dynamically young and
contain multiple \hi~clouds hosting active star forming regions. Thus
care is needed in interpreting the highest $b$-values; even though in
each case the brightest star forming regions are associated with the
potential minimum, observers could expect to observe significant star
formation closer to the GRB sky coordinates than these extreme values
would at first suggest. In addition, small halos have such slow star formation
that they may be invisible for practical purposes.

Furthermore, some caution should be exercised in interpreting
$b$-values smaller than a kiloparsec, since these lie below the
gravitational softening scale of the coarsest simulation included
(Table~\ref{tab:simulations}). Reassuringly, however, we found that
the only effect of removing the low resolution C50 box from our sample
was to truncate the $b$-value distribution above approximately $20
\kpc$, simply reflecting the removal of all high mass halos from our
sample: no change in the low-$b$ tail was produced.

To compare the gas distribution along the sightline of intrinsic and
intervening absorbers, we define $\Delta$ to be the distance over
which the central half of the \hi~column is assembled (i.e. the
physical distance between lower and upper quartile positions). The
distribution of this quantity is overplotted on Figure
\ref{fig:impact} as thin lines (solid for GRBs, dashed for QSOs).
$\Delta$ values suffer equally from the resolution caveat mentioned
for $b$ above; however we can nonetheless conclude that GRB-DLAs are
typically assembled over a longitudinal distance somewhat less than a
kiloparsec whereas QSO-DLAs are built up over larger distances ($\sim
2 \kpc$). This indicates that the \hi~columns in our typical GRB-DLAs
are not assembled from the immediate environment of the GRB itself,
but rather from gas spread through kiloparsec-sized surrounding
regions.

Although the distributions of $b$ and $\Delta$ are seen to be similar
to each other, we found that for individual absorbers these two
quantities are in fact almost uncorrelated; i.e. there is no
significant relation between the longitudinal dimensions of an
absorber and its proximity to the central regions of the host protogalaxy.

%Little conclusion, removed for concisision
%
%Thus typical GRB-DLAs are expected to probe different regions from
%typical QSO-DLAs, most likely arising closer to the centre of the host
%protogalaxy. Our results reinforce such a distinction, but further
%suggest that the typical GRB-DLA-hosting halo is considerably (around
%ten times) more massive than a QSO-DLA-hosting halo.

\subsection{Column density distribution}\label{sec:column-dens-distr}

\begin{figure}
\includegraphics[width=0.5\textwidth]{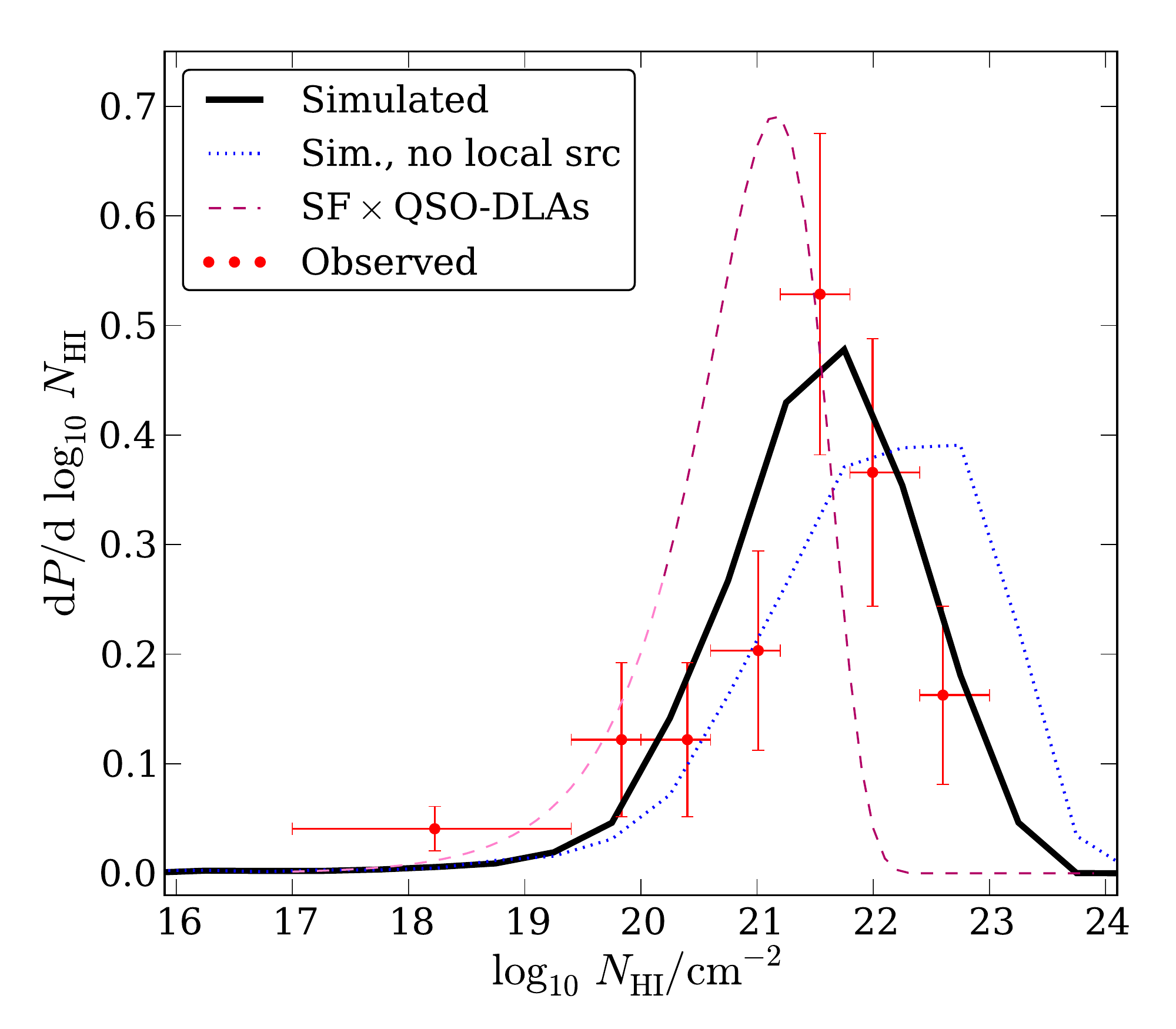}
\caption{The differential column density distribution of GRBs as
  observed (points with error bars; see text for references) and
  simulated (solid line); these are in agreement except for systems
  with $\nhi<10^{19} \cm^{-2}$. The dotted line shows the results
  obtained if stellar ionizing sources are ignored, leading to a
  slight overestimate of very high column density systems. The dashed
  line shows the results expected from a naive argument based on the
  observed QSO-DLA distribution combined with the Kennicutt star
  formation relation; this underestimates the incidence of high column
  density systems, which may be a useful result for understanding the
  high redshift ISM (see text for
  details).}\label{fig:colden}
\end{figure}

%\begin{figure}
%\includegraphics[width=0.5\textwidth]{colden_cumu.pdf}
%\caption{The simulated distributions (thick solid and dotted lines) of
%  Figure \ref{fig:colden} shown cumulatively, highlighting a
%  significant shortfall in low column density systems compared to the
%  observed distribution (stepped solid line). The shaded band around
%  the stepped observed line shows Poisson uncertainties, but these are
%  strongly correlated; the overall significance of the discrepancy for
%  $\nhi>10^{19} \cm^{-2}$ is minimal, as shown in Figure
%  \ref{fig:colden}.}\label{fig:colden-cumu}
%\end{figure}

For systems at $z>1.6$ the Ly$\alpha$ absorption line is redshifted
away from otherwise inaccessible UV wavelengths. For such systems, it
is possible to determine observationally the column density of neutral
hydrogen. We compiled a list of $\nhi$-values for 38 GRB-DLAs in the
redshift interval $2<z<4$ by combining data from
\cite{2006A&A...460L..13J}, \cite{2007ApJ...667L.125C} and
\cite{2009arXiv0907.3449F}. We caution that, unlike for the QSO-DLAs
for which homogeneous catalogues are available
\citep[e.g.][]{2005ApJ...635..123P}, our GRB-DLA statistics are formed
from a heterogeneous sample. It is possible that extinction of
UV/optical afterglows leads to some preference, in the observed data,
to select lower column density absorbers (with correspondingly smaller
dust columns; e.g. \citeauthor{2009arXiv0907.3449F} 2009); at this
stage, however, the observational significance of this effect is not
transparent -- further discussion is given in Section
\ref{sec:model-uncertainties}.

Figure \ref{fig:colden} compares the observed column density
distribution (points with error bars) with our simulated GRB-DLA
column densities (thick solid line). The form of the differential
distribution is well matched by our simulations for $\nhi >
10^{19}\,\cm^{-2}$. It is immediately clear that the GRB $\nhi$
distribution is very different from the intervening case: for
instance, the $\nhi$ distribution diverges at low $\nhi$ for
intervening DLAs but approaches zero for GRB-DLAs; further the median
GRB-DLA has a column density of $\nhi \simeq 10^{21.5} \cm^{-2}$
whereas only $1\%$ of QSO-DLAs are so extreme in a catalogue derived
from SDSS DR5 \citep{2005ApJ...635..123P}\footnote{The DR5 catalogue
  is available from {\tt
    www.ucolick.org/\~{}xavier/SDSSDLA/}. Recently the DR7 catalogue
  was analysed with an independent pipeline
  \protect\citep{2009A&A...505.1087N}; the differences are, however,
  minor for our present purposes.}. We will discuss these
differences and their physical origins in Section
\ref{sec:high-column-dens}.

Overall, the simulated column density distribution is in good
agreement with the observed result. The median $\nhi$ value ($\nhi
\simeq 10^{21.5} \, \cm^{-2}$) is reproduced quantitatively.  The
dotted line in Figure \ref{fig:colden} shows the output from our
simulations without accounting for the effects of local sources
(Section \ref{sec:radi-trans}); in this case a factor two
overabundance of very high column density sightlines
($\nhi>10^{22.5}\,\cm^{-2}$) is predicted. Our simple approach to
accounting for the stellar UV emission has therefore mitigated this
problem. However, our simulations produce a statistically significant
shortfall of systems with $\nhi<10^{19}\,\cm^{-2}$ compared with the
observed population; we will now discuss an estimate of the escape
fraction from our simulations, which will highlight this difficulty.

%Figure \ref{fig:colden-cumu}, which plots the cumulative
%fraction $p(<\nhi)$ from the real and simulated populations,
%highlights this discrepancy. (We caution, however, that care must be
%taken in interpreting cumulative plots, since the errors are strongly
%correlated between different $\nhi$ values. In this case there is a
%misleading apparent discrepancy for $10^{19}<\nhi/\cm^{-2}<10^{21}$
%which is not borne out by examination of Figure \ref{fig:colden}).

\subsection{Escape fraction}

Since we have constructed a sample of neutral hydrogen column
densities emerging from star formation regions, we can calculate
the fraction of radiation at the Lyman limit which escapes
relative to the total produced:

\begin{equation}
f_{\mathrm{esc}} = \int \dd \nhi \, p(\nhi) \, e^{-\sigma_{\mathrm{L}} \nhi}
\end{equation}
where $\sigma_{\mathrm{L}} = 6.28 \times 10^{-18}\,\cm^{2}$ is the
photoionization cross-section of a hydrogen atom at the Lyman limit.
In effect $f_{\mathrm{esc}}$ obeys almost identically
$f_{\mathrm{esc}}=p(<1.6\times 10^{17}\,\cm^{-2})$, since the
exponential in the integral acts as a step function in $\log \, \nhi$
space; however we evaluate the full integral since it is not a
difficult computation. We ignore the effect of dust; although its rate
of UV photon absorption can be comparable to that of the hydrogen in
certain physical situations, \cite{2008ApJ...672..765G} argue
plausibly that this has little impact on the escape fraction since
most radiation leaks out along a small fraction of extremely low
column density (essentially unobscured) sightlines. 
%This line of
%reasoning ties in well with our shallow cumulative distribution for
%$\nhi<10^{19}\,\cm^{-2}$ (Figure \ref{fig:colden-cumu}) which
%indicates that the remaining $1\%$ of absorbers are spread over a wide
%range of very low column densities.

Our calculated escape fraction, $f_{\mathrm{esc}} = 1.0\%$, is
somewhat smaller than recent estimates have suggested.
In particular, it falls short of the $2\%\pm 0.3\%$
derived by \cite{2007ApJ...667L.125C} from reasonable extrapolation of
the observed GRB-DLA statistics \cite[see
  also][]{2009arXiv0907.3449F}; this is a reflection of our previously
described underestimate of the abundance of absorbers in the
observational bin $10^{17}<\nhi/\cm^{-2}<10^{19}$.  More worryingly,
our derived fraction is very sensitive to assumptions we make about
launching the GRB sightlines; for instance, adding a $200 \pc$
gaussian random offset to our stellar particle loci raises the
fraction to $f_{\mathrm{esc}} = 2\%$. Such an approach might be
justifiable if runaway massive stars are preferentially involved in
GRB events \citep[e.g.][]{2006A&A...454..103H}.

While some adjustments can increase the reported escape fraction,
others cause it to decrease. For instance when we restricted
sightlines to originate from younger stellar populations
($<10^7\,\yr$) than our fiducial technique imposes (Section
\ref{sec:sightline-generation}) we derived an escape fraction of
$0.7\%$. We also implemented a model in which the GRB launch rate is
made proportional to the instantaneous star formation rate as
determined by the gas properties (i.e. taking $p_{\mathrm{GRB}}
\propto \rho^{1.5}$ in regions where star formation is permitted, as
opposed to probing new star particles formed by the simulation,
Section \ref{sec:sightline-generation}). In this case the escape
fraction dropped to just $0.1\%$. These changes appear to be because
our stellar feedback algorithm rapidly decreases the density in
regions around young stars in the live simulation, an effect
underestimated in the fully instantaneous model -- but perhaps
overestimated in our standard approach which allows for excessive
delays between formation and GRB events.
% Removed for being confusing to some people..
%
%\footnote{If the escape
%  fraction is in reality so sensitive to the time delays, observed
%  GRB-DLAs may not trace the overall UV escape fraction as accurately
%  as it would first appear, since UV emitting stars have a variety of
%  main sequence lifetimes. }.

None of our other results -- including the column density distribution
for $\nhi>10^{19} \cm^{-2}$ -- are strongly affected by these changes,
so while the escape fraction seems to highlight shortcomings in our
approximate schemes, we are confident that the majority of our results
are robust. Further discussion of this point is offered in Section
\ref{sec:conc-dis}.

\subsection{Interpretation of the high $\nhi$ distribution}\label{sec:high-column-dens}

As described above, both observed and simulated GRB-DLAs have an
entirely different $\nhi$-distribution from that of QSO-DLAs. To
account for this we should consider the relative importance of the
two major distinctions between these populations (halo mass weighting
and internal sightline positioning; Section
\ref{sec:origins-grb-dlas}).  We find that the difference in the
column density distributions arises chiefly from the GRB's propensity
to probe inner regions of the halo, rather than any difference in the
column density distribution as a function of halo mass. 

We verified this by (unphysically) weighting our QSO-DLA sightlines
using the GRB-DLA halo weights ($w_h$; equation \ref{eq:weight}) --
the result was an increase in the median $\nhi$ of just $0.1\,\dex$,
from $10^{20.5}$ to $10^{20.6} \cm^{-2}$. Conversely, by weighting our
GRB-DLA sightlines using the QSO-DLA halo weights (based on the
\hi~cross-section) we found that the median GRB-DLA column density
decreased by just $0.1\,\dex$. This result was foreshadowed by figure
8 of P08 which shows that the mass of the hosting halo has relatively
little effect on the $\nhi$ value of a typical sightline.

The combination of an observationally-derived fit to the QSO-DLA
column density distribution \cite[e.g.][]{2005ApJ...635..123P} and the
Schmidt-Kennicutt relation between gas surface density and star
formation rate ($\dot{\Sigma}_{\star} \propto N_{\mathrm{HI}}^{n}$,
$n\simeq 1.5$; \citeauthor{1998ApJ...498..541K}
\citeyear{1998ApJ...498..541K}) actually predicts a GRB-DLA column
density distribution. This prediction is independent of the details of
the underlying halo populations, relying only on the QSO absorbers to
trace the unbiased neutral hydrogen distribution. If the column
density distribution follows a Schechter function
\citep{1995ApJ...454...69P,2005ApJ...635..123P} with asymptotic slope
$\alpha$ and exponential scale $N_{\gamma}$ one has the following
result for the distribution of GRB $\nhi$ values:
\begin{equation}
\frac{\dd P}{\dd \log_{10} N_{\mathrm{HI}}} \propto N_{\mathrm{HI}}^{1+\alpha+n} \exp\left(\frac{-2 N_{\mathrm{HI}}}{N_{\gamma}}\right)
\end{equation}
where the $1$ in the exponent arises from the change of variables
$N_{\mathrm{HI}} \to \log_{10} N_{\mathrm{HI}}$ and the $2$ in the
exponential arises because, by symmetry, a sightline to the GRB
intersects on average half of the total gas column density of the host
galaxy.

We exhibit this function as a dashed line in Figure \ref{fig:colden},
using the peak posterior dust-corrected\footnote{The dust correction
  is small and does not in fact have a qualitative impact on the
  following considerations, but we include it to remove suspicion that
  dust biasing of QSO samples could account for any discrepancies.}
parameters \citep{PontzenDLADust} determined for the previously
mentioned SDSS DR5 sample in the redshift interval $2.2<z<4$:
$\alpha=-1.79$, $N_{\gamma}=10^{21.61}\, \cm^{-2}$ .  Qualitatively,
this approach predicts the correct form for the GRB-DLA $\nhi$
distribution, with a peak and long tail to low column
densities. However, quantitative details at high $\nhi$ are in some
tension with the observed GRB-DLA results. (One is less interested in
pursuing exact agreement below $\nhi<10^{20.3} \cm^{-2}$ since such
systems are not included in the fit to the quasar data.)

P08 showed that our simulated QSO-DLA column density distribution,
while in good overall agreement with the observed distribution,
over-predicts the occurrence of high $\nhi$ quasar systems. Yet the
present considerations show that the {\it observed} QSO-DLA
distribution {\it under}-predicts the occurrence of high $\nhi$ GRB
systems -- while the simulations are in better agreement with
observations!  This three-way tension between the observed QSO-DLAs,
observed GRB-DLAs and simulations may be a fruitful line of future
exploration to understand details of the high redshift ISM. For
instance, high-$\nhi$ quasar absorbers are plausibly limited by H$_2$
formation \citep{2001ApJ...562L..95S}. The relative over-abundance of
high-$\nhi$ absorbers in GRB afterglows might then suggest
dissociation of molecular clouds in the vicinity of UV-bright young
stellar populations responsible for the GRB events.  Given the
crudeness of our gas phase modelling, we are not in a position to make
any certain conclusions at this stage.

We should take a moment to consider whether it is appropriate to
apply the Kennicutt relation in the sense above. In a direct test of the the relation's
applicability to quasar DLA populations, \cite{2006ApJ...652..981W}
used the $\nhi$ distribution to predict a background level of faint
diffuse emission at rest-frame wavelengths $\lambda \simeq
1500\,\angst$.  After blind searches in deep imaging, they concluded
that the Kennicutt relation was over-estimating star formation in
DLAs. However the authors also noted that this result may be negated
if a substantial fraction of DLAs are associated with nearby bright
compact sources, which were excluded from their search
algorithms. Since this association indeed appears to hold in
our simulations (P08), the relation may tentatively be applied.  We
note that the naive Kennicutt weighting applied to the simulated
QSO-DLAs does produce the approximate distribution of simulated
GRB-DLAs, so this method cannot constitute an entirely spurious
comparison. \cite{2008ApJ...686L..57N} also placed confidence in such
a weighting, using it to adapt their simulated QSO-DLA statistics to
discuss the expected redshift evolution of the GRB-DLA $\nhi$
distribution.

\subsection{Metallicity Distribution}\label{sec:metall-distr}

\begin{figure}
\includegraphics[width=0.5\textwidth]{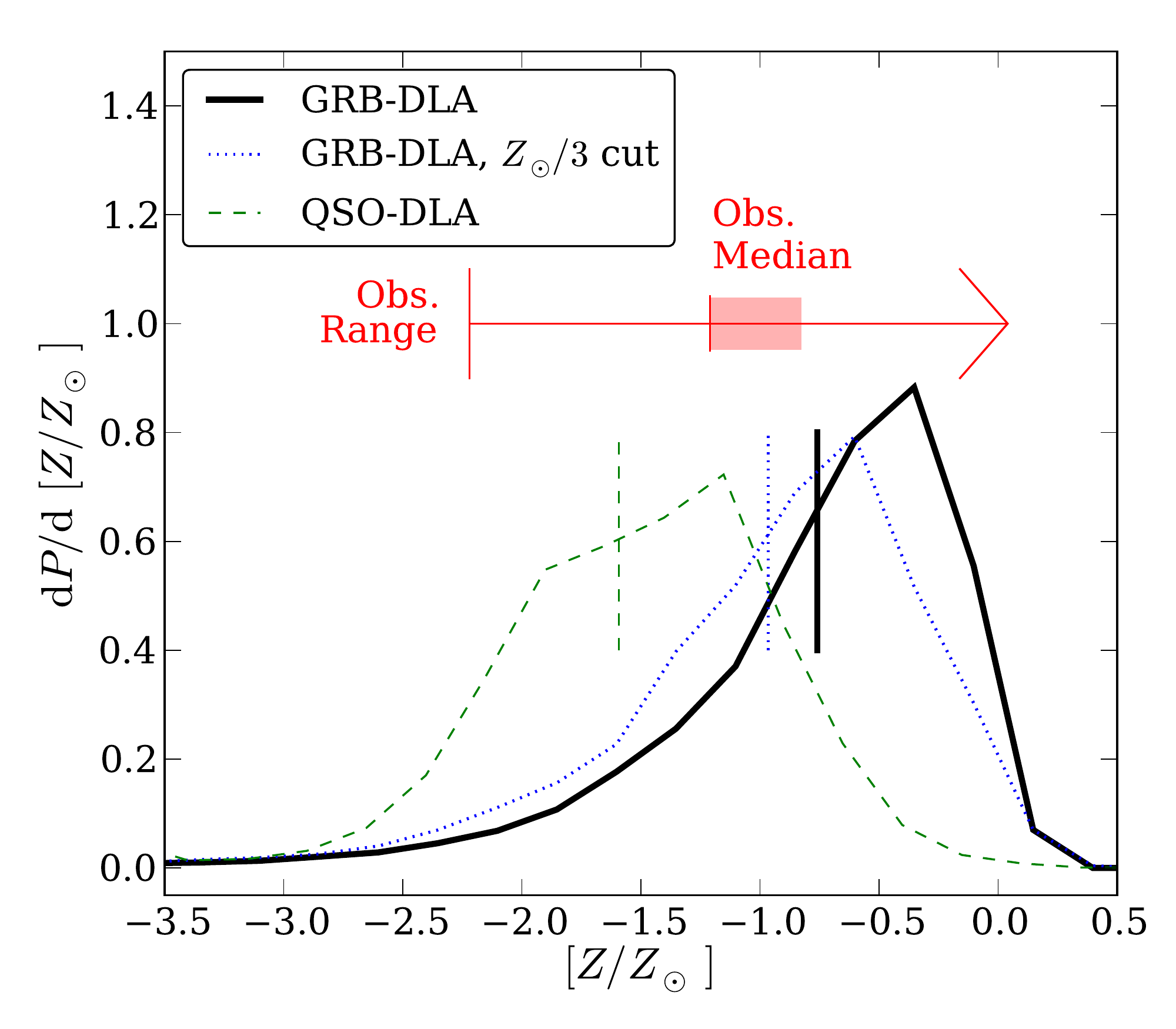}
\caption{The neutral gas metallicity of our simulated GRB-DLA
  sightlines (solid line) compared with the same diagnostic of QSO-DLA
  sightlines (dashed line). GRB-DLAs have significantly higher
  metallicities as they are associated with star forming regions of
  more massive halos than QSO-DLAs. This picture is broadly supported
  by the handful of measurements of (and lower limits on) GRB neutral
  gas metallicities at $z>2$, the range of which is shown by the
  horizontal bar.  The shaded region shows the range of the median if
  we assume systems with line saturation have up to three times the
  metallicity of their measured lower limits (see text for
  details). The dotted line shows the simulated results after imposing
  a restriction that GRBs can only arise in stellar progenitors with
  metallicity $Z<Z_{\odot}/3$.} \label{fig:metal}
\end{figure}

Each of our sightlines is assigned a neutral gas metallicity according
to equation (\ref{eq:sl-metal}); using our fiducial techniques we can
then predict the metallicities observed in cosmological GRB-DLA
samples. Figure \ref{fig:metal} shows the resulting distribution as a
solid thick line, with the median indicated by a short vertical line
bisecting the distribution. Our metallicity distribution peaks at
around $Z_{\odot}/3$, with a sharp decline to higher metallicities
(and long tail to lower); as a result the median is approximately
$Z_{\odot}/6$. 

As a rough observational comparison, we used the lists
of \cite{2007ApJ...666..267P,2009ApJ...691L..27P} and \cite{2009arXiv0907.1057L}
to compile eight GRB-DLA metallicity measurements and a further seven lower limits
at $z>2$; to our knowledge this is a complete census of such constraints.
The small numbers, line saturation effects and unknown
biases\footnote{If anything, the observed sample is likely biased in
  favour of low metallicity GRB-DLAs which have smaller dust columns
  and therefore brighter optical afterglows; see
  e.g. \protect\cite{2008ApJ...683..321F}.}  make it hard, at present,
to attempt a quantitative comparison of the observed and simulated
metallicity distributions.  We merely indicate in Figure
\ref{fig:metal} the range of the observed metallicities (by a
horizontal bar), with a vertical line indicating the median value.
The shaded region shows the range of the median if we include assumed
metallicity measurements of up to three times the value of the lower
limits.

Given the uncertainties our simulations seem to provide an acceptable
match. Although the observational constraints are not tight, there is
no doubt that, for instance, our QSO-DLA metallicity distribution
(shown as a dashed line with its median similarly indicated) would be
deemed inconsistent with the data\footnote{Our new simulations have a
  QSO-DLA metallicity distribution with a median $[Z/Z_{\odot}]\simeq
  -1.6$, approximately $0.2 \, \dex$ lower than the median in P08, as
  described in Section \ref{sec:techniques}.}.

Unlike for the column density distribution, we attribute much of this
metallicity shift to the $1 \, \dex$ higher halo masses probed by the
typical GRB-DLA compared with a QSO-DLA (Section
\ref{sec:origins-grb-dlas}), in agreement with \cite{2008ApJ...683..321F}.  
Mean neutral gas metallicity in our
halos range from $\sim 10^{-3} Z_{\odot}$ for $M_{\mathrm{vir}}=10^9
M_{\odot}$ to $\sim 10^{-1} Z_{\odot}$ for $M_{\mathrm{vir}} = 10^{12}
M_{\odot}$: a $0.7 \dex$ rise is expected from this effect. The
remainder of the effect is accounted for by
the marginally higher metallicities of the central regions which the
GRB sightlines probe. This is a small $\sim 0.1 \dex$ correction,
despite large internal
abundance gradients discussed below, because the sightline averaging
smears out much of the  metal density variations.

\subsection{Effect of GRB Metallicity Ceiling}\label{sec:metal-ceiling}

\begin{figure}
\includegraphics[width=0.5\textwidth]{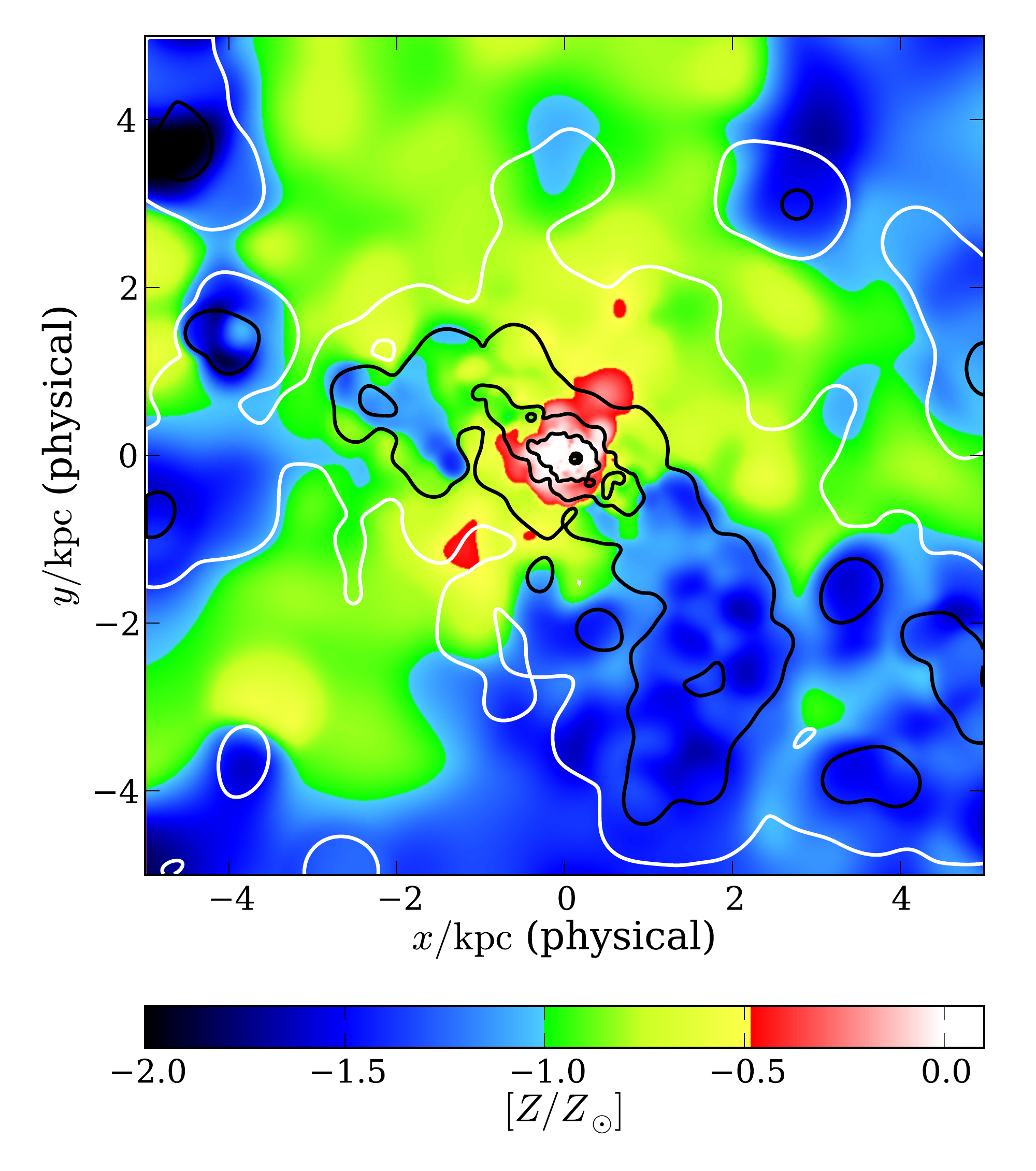}
\caption{Map of the metallicity of the gas in a thin slice through the
  centre of the most massive halo in our MG box ($\mvir \simeq 3
  \times 10^{10} \msol$), with the volume density of \hi~overplotted
  as contours (the outermost white contour corresponds to
  $n_{\mathrm{HI}} =\, 10^{-2} \cm^{-3}$ and inner, black, contours
  indicate $0.1, 1, 10$ and $10^2\,\cm^{-3}$). Little of the gas by
  volume exceeds $1/3$ solar (red region, the cut-off in our truncated
  GRB model), and star formation in nearby regions can still give rise
  to GRBs with lines-of-sight intersecting the high $Z$
  region.} \label{fig:metalmap}
\end{figure}

So far we have assumed that GRB events simply trace, in an unbiased
fashion, star formation activity in the cosmos. However, GRBs are
thought to arise from rapidly rotating progenitor stars.  Since
angular momentum is concentrated in outer layers of high mass stars,
it is easily lost through winds.  Mass loss rates are thought to scale
almost linearly with metallicity, so GRBs could well arise
preferentially from low metallicity progenitor stellar populations
\citep[][and references therein]{2006ARA&A..44..507W}. There is also
some observational evidence, particularly at low redshifts, that
GRB-hosting populations are indeed biased in this sense
(e.g. \citeauthor{2006Natur.441..463F} 2006,
\citeauthor{2008AJ....135.1136M} 2008,
\citeauthor{2009ApJ...693.1236C} 2009; although see
\citeauthor{2007MNRAS.375.1049W} 2007, 
\citeauthor{2008ApJ...687.1201K} 2008,
\citeauthor{2009ApJ...691..182S} 2009).

We tested the effect of such a picture on our results by producing a
new set of sightlines, allowing GRBs to be launched only from stellar
populations with $Z<Z_{\odot}/3$ \citep[as suggested by][]{2006ApJ...637..914W} 
and re-calculating the halo weights
according to $w_h' = w_h f(Z<Z_{\odot}/3)$ where $w_h$ is defined by
(\ref{eq:weight}) and $f(Z<Z_{\odot}/3)$ is the mass fraction of young
stars (according to our criterion in Section
\ref{sec:sightline-generation}) with metallicity below the adopted
limit.

We found that the differences in column density and impact parameter
distributions are entirely negligible; the only significant effect was
seen in the sightline neutral gas metallicities and the stellar
population metallicities (the latter quantity is discussed below in
Section \ref{sec:grb-host-halo}). The updated neutral gas metallicity
distribution is shown as a dotted line in Figure \ref{fig:metal} and
shows a reduction in median metallicity by $0.2\,\dex$ to
$[Z/Z_{\odot}]=-0.9$. This is a relatively small shift, largely
reflecting that even our most massive halos at $z=3$ have mean cold
gas metallicities of only $Z_{\odot}/2$.

Furthermore sightlines of up to solar metallicity remain attainable
with the cut because the neutral gas within each halo has metallicity
dispersions of up to an order of magnitude; thus sightlines which
start in a low metallicity environment may nonetheless encounter a
nearby dense knot of solar metallicity gas which dominates the
line-of-sight measurement. This is illustrated by Figure
\ref{fig:metalmap}, which shows the metallicity of the gas in a thin
slice through the centre of the largest halo in the MG box. Red and
white regions have metallicity such that they are excluded from
hosting GRB progenitors. Overplotted contours show the regions in
which the neutral gas volume density exceeds
$n_{\mathrm{HI}}=10^{-2}$, $0.1$, $\cdots$, $10^2\,\cm^{-3}$
(respectively from outer to innermost contour). Dense neutral regions
($>0.1 \cm^{-3}$) have mean metallicity $10^{-0.4}\,Z_{\odot}$ with a
dispersion of $0.8\,\dex$. Qualitatively GRBs are to be found in the
densest regions which are not forbidden by the metallicity cut (i.e.
not within the red area); the prime location remains very close to the
centre so that high metallicity gas will present a large cross-section
to the sightline.

\subsection{GRB host halo properties}\label{sec:grb-host-halo}

\begin{figure}
\includegraphics[width=0.5\textwidth]{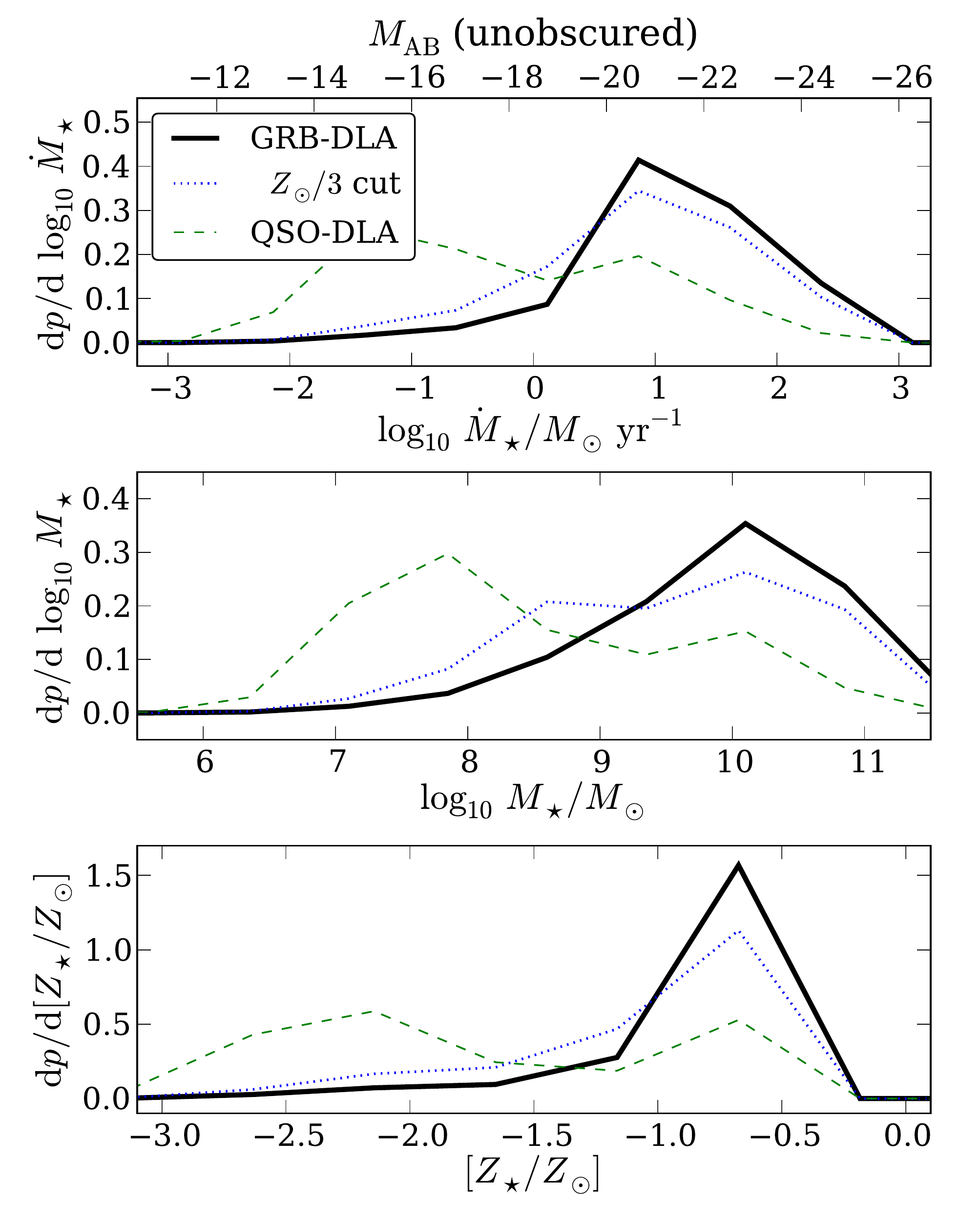}
\caption{Properties of the stellar population within the host galaxies
  of $z=3$ GRB (thick solid line, dotted line for $<Z_{\odot}/3$
  progenitors) and QSO-DLAs (dashed line). From top to bottom we plot
  the probability densities distributions for star formation rate,
  total stellar mass and (mass-weighted) stellar population
  metallicity respectively. A second scale on the uppermost panel
converts the star formation rate to
  an ultraviolet AB magnitude as described in the text, ignoring dust
  obscuration. The properties of the host galaxies of the two classes
  of absorber differ in accordance with their different virial masses,
  with the GRBs favouring somewhat higher star formation rates, total
  stellar masses and stellar metallicities. This applies even for the
  cut metallicity GRB population, although the high-$Z$ peak is
  slightly suppressed in this case.} \label{fig:galprops}
\end{figure}

Our simulation naturally associates each QSO- or GRB- sightline with a
host halo, the properties of which can readily be computed. In this
section we briefly describe the distribution of star formation rates,
total stellar masses and stellar metallicities in the populations
within each halo. 

Figure \ref{fig:galprops} shows the distribution of each of these
properties (in three separate panels) with the GRB-DLAs and QSO-DLAs
represented by solid and dashed lines respectively. The dotted lines
show the population of GRB-DLAs if a progenitor metallicity criterion
$Z<Z_{\odot}/3$ is introduced (see Section \ref{sec:metal-ceiling}).
The GRB-DLA hosts generally show somewhat higher star formation rates,
total stellar masses and metallicities than the intervening
population: mean star formation rates\footnote{Calculated as described
  in P08 section 4.3. The QSO distribution shown in the middle panel
  of Figure \ref{fig:galprops} is equivalent to figure 14 of P08;
  however the latter plot shows $\dd p/\dd \dot{M}_{\star}$, not $\dd
  p/\dd \log_{10} \dot{M}_{\star}$ as its label in the published
  version erroneously claims -- this was a typographical error.} for
the GRB and QSO DLAs are $10\, \msol\,\yr^{-1}$ and
$1\,\msol\,\yr^{-1}$, a difference of $1\, \dex$. However this should
be compared with a wide dispersion of around $4\, \dex$ in both cases.
We have added to the top panel a scale showing corresponding
unobscured UV magnitudes using the luminosity conversion ratio from
\cite{1998ARA&A..36..189K}, corrected by a factor $1.6$ in order to
consistently use the Kroupa IMF assumed by our simulations.

Mean stellar masses for the QSO-DLA and GRB-DLA host populations
differ by around an order of magnitude (with respective values
$10^{8.9}\, \msol$ and $10^{9.9}\,\msol$), but exhibit wide
dispersions around these values. For the stellar metallicity (distinct
from the neutral gas metallicities described in Section
\ref{sec:metall-distr}) the mean values are $10^{-0.7}\,Z_{\odot}$,
$10^{-0.9}\,Z_{\odot}$ for unbiased and metallicity-biased GRBs
respectively compared against $10^{-1.4}\,Z_{\odot}$ for the QSO DLAs.

The agreement of these results with observational constraints have
been discussed elsewhere for QSOs (P08). For GRBs, no homogeneous high
redshift catalogues of host galaxies have been assembled. However the
four $z\gsim 1$ objects studied by \cite{2004A&A...425..913C} have
star formation rates well within the range suggested by our
simulations; see also \cite{2005MNRAS.362..245J}.  At $z>2$, the eight
GRB-DLAs with identified host galaxies listed by
\cite{2006Natur.441..463F} have a range of luminosities $-16\ge M_V
\ge -22$ in the observer-frame V band; broadly, given that our
brightest objects should be substantially dust obscured in the
rest-frame UV, our results are consistent with these observations.

All of the distributions can be interpreted by inspecting the
relation between virial mass and the property under consideration
(see, for example, figure 13 of P08 for the star formation rate).  As
well as satisfactorily accounting for the approximate means in each
case, specific features in the distribution function can be explained
in this way.  In particular $Z(\mvir)$ rises steeply from $10^{-3}
Z_{\odot}$ at $\mvir = 10^9 \,\msol$ to around half solar at $10^{11}
\msol$, beyond which the metallicity remains constant. Physically this
is because massive halos harbour older stellar
populations as well as the young enriched stars from recent formation
activity. This gives rise to bimodality in the metallicity of the
QSO-DLA stellar populations, with peaks at $Z \sim 10^{-2.0} \,
Z_{\odot}$ (corresponding to the metallicity of the peak QSO-DLA mass,
$10^{10}\, \msol$) and $Z \sim 10^{-0.5}\,Z_{\odot}$ (corresponding to
the plateau metallicity: all halos with $M>10^{11} \,\msol$ fall into
this bin). Bimodality is not seen in the neutral gas metallicities
because these measurements are averaged only along the line of sight,
not the entire halo, and local metallicities can reach values of up to
solar: see Figure~\ref{fig:metalmap}.

\section{Conclusions and Discussion}\label{sec:conc-dis}

\subsection{Overview}

We have investigated the properties of high redshift ($z \sim 3$)
GRB-DLA absorbers in a series of galaxy formation simulations, on the
assumption that long duration GRBs are associated with the death of
massive, short-lived stars. Following our success in matching
properties of traditional DLAs along randomly chosen sightlines to
quasars (P08), we used our results to interpret observational
differences between the populations, finding two primary distinctions.

Firstly, because the star formation rate scales more steeply with
virial mass ($\dot{M}_{\star} \sim \mvir^{1.6}$, P08 figure 13) than
does the neutral hydrogen cross-section (approximately
$\sigma_{\mathrm{DLA}} \sim M_{\mathrm{vir}}$, although a single
power-law is an oversimplification; P08 figure 4), the halos
contributing most to the GRB-DLA incidence rate are considerably more
massive than those contributing to the QSO-DLA incidence rate (Section
\ref{sec:origins-grb-dlas}). The dominant contribution for GRB-DLAs
arises in halos with $10^{10}<M_{\mathrm{vir}}/M_{\odot}<10^{12}$; for
QSO-DLAs this range is an order of magnitude lower,
$10^{9}<M_{\mathrm{vir}}/M_{\odot}<10^{11}$. While this argument
obviously relies on our simulated scalings being accurate, there are
many reasons for being cautiously optimistic: in particular, the
simultaneous matching of the QSO-DLA metallicity distribution (P08)
and the galaxy mass-metallicity relation \citep{2007ApJ...655L..17B}
is hard to reproduce unless the underlying relations are
realistic. \cite{2008ApJ...683..321F} have previously reported that
GRB metallicities can be accounted for if a Holmberg-like relation,
$\sigma_{\mathrm{DLA}} \propto \dot{M}_{\star}^{2t}$ with $t\simeq
0.4$, holds at high redshift; here we have shown that the
metallicities are reproduced and that just such a scaling arises
naturally from cosmological simulations (albeit with $t \sim
0.3$). The difference in mass ranges has some effect on the predicted
properties of the hosting halos, most notably lifting the mean stellar
metallicity from $10^{-1.4}\,Z_{\odot}$ for QSO-DLAs to
$10^{-0.7}\,Z_{\odot}$ for GRB-DLAs. However, these effects are
relatively small compared to the large spread of hosting halo
properties in both populations, reflecting that incidence rates for
both classes are a shallow function of virial mass.

\cite{2004MNRAS.354..581C} took a reciprocal approach, using observed
galaxy properties to determine the likely underlying population of
host objects. Their results showed that GRB hosts at $z=3$ have
typical masses $M_{\mathrm{vir}}<10^{10} \msol$ which does not agree
with our analysis. This may be because \cite{2004MNRAS.354..581C}
adopt, on an observational basis, the star formation efficiency (rate
of star formation per unit stellar mass) as an indicator of GRB
probability. Our simulations do not suggest a strong link between star
formation efficiency defined in this way and the GRB event rate; if
extended observational analysis continues to point to such a link,
further investigation will be required.

The second distinguishing effect we find is that the GRB-DLAs arise in
gas closer to the centres of the hosting objects than a conventional
DLA \cite[as seen observationally, e.g.][]{2003ApJ...585..638S,2004A&A...419..927V,2006Natur.441..463F}. While resolution
artifacts cloud an exact interpretation of some details, we are
confident in concluding that approximately $50\%$ of GRB-DLAs arise
less than $1 \kpc$ from the centre of their parent halo, whereas this
figure for QSO-DLAs is just $10\%$ (and, accordingly, the QSO-DLAs
have a higher median impact parameter of $\sim 4 \kpc$). Similarly,
GRB-DLAs tend to be more concentrated along the line of sight; we
defined the measure $\Delta$ to be the distance over which half the
total hydrogen column density is accumulated and found that this took
a median value of $1 \kpc$ for GRB-DLAs and $2 \kpc$ for QSO-DLAs.
This indicates that the neutral hydrogen seen in GRB afterglow spectra
is not to be found in the immediate circumstellar environment, but is
spread over an extended region -- although a detailed analysis would
require better resolution of the ISM than our present simulations are
able to provide.

\subsection{Observational diagnostics}

We investigated the neutral hydrogen column density (Sections~
\ref{sec:column-dens-distr},~\ref{sec:high-column-dens}) and the
metallicity (Section \ref{sec:metall-distr}) distributions of our
simulated absorbers, finding good observational agreement in both
cases except for a deficiency in low column density ($\nhi<10^{19}
\cm^{-2}$) absorbers.  
The simulated neutral
hydrogen column densities, as for the observed GRB absorbers, are much
higher than those seen in intervening DLAs. We conclude that this is
largely due to the lowered impact parameters, with the change in the
masses of halos having a secondary effect (for instance, figure 8 of
P08 shows that the relation between column density and halo mass is
relatively weak).

The metallicities are higher, by almost a factor of ten, than those
seen in QSO-DLAs -- even if GRB-DLAs can only arise from stellar
populations with $Z<Z_{\odot}/3$ \cite[a limit motivated by the
  collapsar picture, e.g.][]{2005A&A...443..581H,2006ApJ...637..914W}.
This suggests that the observed population of high redshift GRB-DLAs
and GRB hosts
do not pose any severe problems for the typical collapsar GRB
paradigms, counter to the conclusions of \cite{2007MNRAS.375.1049W}
and \cite{2008MNRAS.386..608L}, but in agreement with
\cite{2007MNRAS.375..665N}. The ineffectiveness of a metallicity
ceiling on the GRB-DLA progenitors to limit the metallicity of the
overall sightline population arises because, within each protogalaxy,
neutral gas metallicity dispersions of up to an order of magnitude are
possible (greatly exceeding the $0.2\, \dex$ allowed by
\citeauthor{2007MNRAS.375.1049W} 2007). In our picture GRBs
now originate in dense regions with lower metallicities, but they are
typically still close to gas with significantly higher metallicity
(Figure \ref{fig:metalmap}) which consequently presents a large
covering fraction. Furthermore high mass, metal rich halos continue to
contribute significantly to the GRB rate: even for our most massive
galaxy at $z=3$ (with a mean neutral gas metallicty of $\simeq
Z_{\odot}/2$), $30\%$ of the star-forming gas mass remains at low
metallicities $Z<Z_{\odot}/3$.  We note that, since our simulations
include a turbulent metal diffusion term, this dispersion cannot be
attributed solely to numerical effects (Section~\ref{sec:techniques}).

The leaves open the question of whether the observational situation at
lower redshifts ($z\lsim 1.5$), where host galaxies may be less
massive than typical hosts of supernovae
(\citeauthor{2006Natur.441..463F} 2006; although see
\citeauthor{2008ApJ...687.1201K} 2008,
\citeauthor{2009ApJ...691..182S} 2009), can be quantitatively
reconciled with less pronounced differences between supernovae and GRB
hosts at high redshift. A natural explanation is simply that at
earlier times mean metallicities are lower and high metallicity gas is
typically less well mixed
\citep{2007MNRAS.375..665N,2009arXiv0905.1953K}. In principle we could
probe this claim in our simulations, but such work is beyond the scope
of the present investigation.

\subsection{UV Escape fraction}

One of the shortcomings of our work is the unreliability of our
predicted $1\%$ escape fraction of UV photons from gamma ray burst
regions.  This is somewhat lower than recent simulations of escape
fractions ($\sim 2\%$ from results by \citeauthor{2008ApJ...672..765G}
2008; between $2\%$ and $10\%$ according to
\citeauthor{2006ApJ...651L..89R} 2006; or, for halos with masses
$M\lsim 10^9 \msol$, \citeauthor{2009ApJ...693..984W} 2009 reported
escape fractions between $25\%$ and $80\%$ in very high resolution
Eulerian simulations). Furthermore it falls short of constraints on
escape fractions from comparing rest-frame UV emission on either side
of the Lyman limit break \citep[$\sim 2\%$, although this value is
  sensitive to assumptions about the dust
  extinction;][]{2006ApJ...651..688S} and most directly underpredicts
the escape fraction implied by current GRB observations \citep[$\sim
  2\%$;][also visible directly in Figure
  \ref{fig:colden}]{2007ApJ...667L.125C}.  However, most importantly,
it is not robust to changes in our sightline construction algorithms
(Section \ref{sec:column-dens-distr}), unlike our other reported
results.

The fragility reflects enormous uncertainties in this regime.  In
addition to the inherent complexities in accurately simulating
absorption systems, for GRB-DLAs one can no longer justify the
assumption that local ionizing UV sources are unimportant compared
with cosmological sources (Section \ref{sec:radi-trans}). We have
taken a simple pragmatic approach to this problem by noting that the
ionizing radiation of local sources is largely absorbed by
\hii~regions which we paint into our output.  (More accurate radiative
transfer would probably lend confidence to our results, but we caution
that without properly resolved \hii~regions the physical meaning of
even the most precise transfer algorithms is still obscure.)
Furthermore, there is a sensitivity to the energy injection from
supernovae, as demonstrated by the drop in escape fraction to $0.1\%$
when triggering GRBs based on the star formation rate scaling of our
gas particles (rather than existing young stellar particles which will
have injected feedback energy into the nearby ISM). Finally, if
runaway stars are important contributors to the GRB incidence rate
\citep{2006A&A...454..103H} one could expect a significant increase in
the number of low column density absorbers.

\subsection{Model Uncertainties}\label{sec:model-uncertainties}

There are further complications specific to GRB-DLAs which we have not
specifically addressed. One is the evidence, from fine-structure
transitions, that the neutral phase metals detected in the systems are
found at distances ranging from hundreds of parsecs
\citep{2006ApJ...648...95P} to as much as a kiloparsec or more
\citep{2007A&A...468...83V,2009ApJ...694..332D}. It is possible that
the GRB afterglow itself ionizes a significant amount of material
along the line of sight. Given the known afterglow curves, the
ionization of material by the GRB itself has been advanced as a
possible explanation (e.g. \citeauthor{2007A&A...468...83V} 2007),
although whether bubbles of more than a few parsecs radius can be
ionized is unclear (e.g. \citeauthor{2008ApJ...685..344P} 2008). The
afterglow-ionizing scenario is certainly not firmly established:
perhaps the hot progenitor stellar populations have already carved out
large ionized bubbles within the ISM, or the densest knots of gas
which are metal rich are simply found at some distance from the GRB
site (as we have mentioned above). Unfortunately we do not have
resolution to probe the required scales reliably, but we did produce
comparison statistics in which we ionized all material within a $1\,
\kpc$ sphere of a GRB site before measuring its sightline
properties. This led to a decrease in the numbers of very high column
density absorbers ($\nhi>10^{22} \cm^{-2}$), reducing the median to
$\nhi=10^{20.9} \cm^{-2}$. While the agreement of our simulations with
observations was therefore impaired, our results were not
qualitatively altered. Simulations with better resolution of the ISM
will be necessary to properly constrain the effects of GRBs on their
immediate environment.

Similarly we have been unable to make a quantitative assessment of the
possible biasing effects of dust. It is possible that dust obscuration
of optical afterglows significantly biases against observing high
metallicity, high column density absorbers. Indeed the phenomena of
`dark bursts', GRBs for which the optical afterglow is not seen or is
dimmed by several magnitudes, could be explained if these bursts are
simply obscured by large columns of dust
\citep{2009arXiv0905.0001P}. However, it appears to be inappropriate
to extrapolate the dust columns predicted for quasar absorbers
\citep[by models such as those in][]{PontzenDLADust} to our simulated
GRB results: such an approach would significantly over-predict the
reddening seen in real GRB-DLA samples \citep[for a compilation
  see][]{2009arXiv0907.3449F} and would suggest, for example, that GRB
080607 (with $\nhi=10^{22.7}\,\cm^{-2}$, $Z \sim Z_{\odot}$ according
to \citeauthor{2009ApJ...691L..27P} 2009) should have been completely
dark. Either the GRB-DLAs arise in regions in which simple scalings of
dust columns fail, or perhaps the intense outgoing radiation is able
to rapidly destroy dust even beyond the hydrogen ionization
front. Whatever the explanation, we have been unable to make full
sense of this issue in our present exploration and leave its
resolution to future work.

Despite these difficulties, we have affirmed in the present work that
GRB-DLAs offer a new insight into the inner regions of the
protogalactic population at high redshifts. Their basic statistics can
be matched by simple techniques applied to existing cosmological
simulations.  There can be little doubt that, as observational
datasets increase in size and sophistication, GRBs and their
associated absorbers will make a significant contribution to our
understanding of the structure and formation of galaxies.

\vspace{-0.3cm}
\section*{Acknowledgements}

We acknowledge helpful conversations with Jason
X. Prochaska, Susanna Vergani and Lise Christensen. AP is supported by
a STFC studentship and scholarship at St John's College, Cambridge.
FG acknowledges support from grants HST GO-1125, NSF ITR grant
PHY-0205413, NSF AST-0607819 and NASA ATP NNX08AG84G.  We thank the
computer resources and technical support by TERAGRID, ARSC, NAS and
the UW computing center, where the simulations were run.

 \bibliographystyle{mn2e} {\small
  \bibliography{/home/app26/documents/refs}}

\end{document}